\documentclass[twocolumn,showpacs,preprintnumbers,amsmath,amssymb,prb,aps]{revtex4-1}
\usepackage{epsfig}
\usepackage{epstopdf}
\usepackage{graphicx}
\usepackage{dcolumn}
\usepackage{bm}
\usepackage{textcomp}
\usepackage{array}
\usepackage{subcaption}
\usepackage{booktabs}
\usepackage{blkarray} 

\DeclareUnicodeCharacter{2212}{-}
\begin{document}

\title{Evidence of phase stability, topological phonon and temperature-induced topological phase transition in rocksalt SnS and SnSe}

\author{Antik Sihi$^{1,}$}
\altaffiliation{sihiantik10@gmail.com}
\author{Sudhir K. Pandey$^{2,}$}
\altaffiliation{sudhir@iitmandi.ac.in}
\affiliation{$^{1}$School of Basic Sciences, Indian Institute of Technology Mandi, Kamand - 175075, India\\
$^{2}$School of Engineering, Indian Institute of Technology Mandi, Kamand - 175075, India}

\date{\today}

\begin{abstract}
  Both SnS and SnSe have been experimentally and theoretically confirmed as topological crystalline insulators in native rocksalt structure. Here, phononic structure, thermodynamical properties and temperature dependent electron-phonon interaction (EPI) have investigated for both the materials in rocksalt phase. Previously performed theoretical studies have predicted the phase instability of SnS in this crystal structure at ambient condition. But, after a detailed study performing on the phonon calculation of SnS, we have predicted the phase stability of SnS with considering the Sn 4$p$ orbitals as valence states in $ab-initio$ calculation. The importance of long range Coulomb forces along with the themodynamical properties are also described in detailed for both materials. The computed value of Debye temperature ($\Theta_D$) for SnS (SnSe) is $\sim$315.0 K ($\sim$201.7 K). The preliminary evidence of topological phonon is found along X-W direction, where the linear band touching is observed as compared to type II Weyl phononic material ZnSe \cite{liu_weylphn}. The topological phase transition is seen for these materials due to EPI, where non-linear temperature dependent bandgap is estimated. The predicted value of transition temperature for SnS (SnSe) is found to be $\sim$700 K, where after this temperature the non-trivial to trivial topological phase is seen. The strength of EPI shows more stronger impact on the electronic structure of SnS than SnSe material. The reason of non-linear behaviour of bandgap with rise in temperature is discussed with the help of temperature dependent linewidths and lineshifts of conduction band and valence band due to EPI. The present study reveals the phase stability of SnS along with the comparative study of thermal effect on EPI of SnS and SnSe. Further, the possibility of temperature induced topological phase transition provides one of important behaviour to apply these two materials for device making application.   

\end{abstract}

\maketitle
\small

\section{Introduction} 

  Nowadays, topological materials (TM) are conquering modern research of condensed matter physics both in theoretical and experimental studies due to the presence of unique non-trivial behaviour of these materials \cite{hasan_rev,bansil_rev}. In such materials, symmetry always plays an important role in deciding the non-trivial properties. In case of topological crystalline insulators (TCI), the bulk band structure shows insulating behaviour, whereas the surface states are conducting in nature even without the presence of time-reversal symmetry \cite{fuprl}. These surface states for TCI materials are typically protected by mirror or different crystal symmetries \cite{tci_sym1, tci_sym2}. Tin chalcogenides based materials are mostly classified as TCI after theoretical and experimental evidences \cite{hsieh, tanaka, ref_pbsnte, ref_pbsnse}. SnTe is the first theoretically and experimentally evident material in the class of TCI. Moreover, motivated by the above mentioned work, SnS and SnSe are also categorized as TCI by performing theoretical calculations \cite{y_sun,yakovkin}. The further experimental works were carried out by different research groups on SnSe to evident the topological behaviour \cite{wang_snse_exp1,jin_snse_exp2}. Till today, it is known that experimentally rocksalt SnSe is formed in ambient conditions \cite{wang_snse_exp1,jin_snse_exp2,wang_snse}, whereas SnS is synthesized by the epitaxial growth on some proper substrate \cite{mariano_exp_sns, bilenkii_exp_sns2, skelton_2017}. First-principle phonon calculation of any sample provides the information of dynamical stability of the corresponding material \cite{ashcroft}. Therefore, some of phonon calculations were already performed on SnS and SnSe, where it has been predicted that SnSe is possible to form in normal condition, whereas SnS in ambient condition is dynamically unstable \cite{skelton_2017,skelton_2021}. Sk $et.$ $al.$ have showed in their work that the property of phonon dispersion of any material depends on the role of orbitals of corresponding atoms in $ab$-$initio$ calculation for computing the force exerted on the atoms \cite{sk_2021}. The force on the atoms are estimated by separating the total orbitals of particular atom into core and valence for carrying out the calculation. Moreover, proper choices of the core and valence orbitals in density functional theory (DFT) based calculation have provided the fate of phonon dispersion. Therefore, motivated from this work, it will be really interesting to study for SnS whether the similar situation is found for this topologically important material or not. Moreover, it is seen for SnTe that the experimental phonon dispersion shows nice matching with the theoretically calculated phonon branches when the non-analytical term correction (NAC) is added with the phonon calculation \cite{antik_pla}. Typically, the NAC in theoretical calculation provides the effect of long range Coulomb forces for ionic crystal \cite{togo_2015}. This study will also give the information of LO-TO splitting of the corresponding phonon branches of particular material. Therefore, the effect of long range Coulomb forces for these two tin chalcogenides ($i.e.$ SnS and SnSe) are needed to be explored for getting more detailed description of phonon branches. Further, the possibility of topological phase transition always brings more applicability of any topological material. Therefore, the investigation of topological phase transition may reveal some insightful information for these materials.  
  

  The transition between topological phase to normal insulating phase is generally observed when the bandgap closes and the band inversion in valence and conduction bands is removed. Moreover, the valence and conduction bands touch only at the transition temperature point. However, it is noted that above and below the transition temperature, the sample will show semiconducting behaviour. The different external parameters like pressure \cite{pressure_1,pressure_2}, temperature \cite{temp1,temp2, temp3}, defects etc. are typically responsible for the topological phase transition. Temperature induced transition between topological to normal insulating phase for any material always brings more available options for device making. It is known that increment of temperature will directly affect the electron-phonon interaction (EPI) within the material along with electron-electron interaction (EEI). The temperature effect on the EEI of SnS, SnSe and SnTe are discussed in literature, where $GW$ method in Matsubara-time domain is used for considering the temperature effect \cite{antik_jpcm,asihi_sse}. Recently, Querales-Flores $et.$ $al.$ had shown in their theoretical studies that the topological phase transition of rocksalt SnTe material is found due to EPI \cite{flores}. Therefore, detailed investigations of EPI for SnS and SnSe are necessary to carry out for enhancing the usefulness of these two materials in case of practical applications. The transition of topological phase of SnSe is found to be important for making optocaloric cooling device \cite{zhou_opto}. This transition is also expected to show different transport properties in normal insulating phase than topological phase \cite{wang_thermo_rs}. Also, understanding the intrinsic scattering like EPI of any material is important for tailoring the thermoelectric properties of any particular material \cite{shiva_jpcm_sctter}.    
  
  Typically, first principle and model Hamiltonian based formalism are used to encounter the study of temperature induced topological phase transition through EPI. Nowadays, it is well known practice to calculate the electron-phonon matrix elements by using density functional perturbation theory (DFPT) with the help of Wannier functions for interpolation method \cite{giustino_wann1}. The use of Wannier function reduces the computational cost to compute the electron-phonon matrix elements than the usual ways. The changes in bandgap due to EPI and the estimation of self-energy for electrons and phonons are also possible to get from this calculation. The information of self-energy provides the insight of many-body interaction effect of any material, which is useful for understanding the interaction in depth. This method is expected to provide the changes of electronic states due to thermal effect with properly estimating the EPI for typical topological material \cite{flores}.      

\begin{figure*}
  \begin{center}
    \includegraphics[width=0.9\linewidth, height=10.5cm]{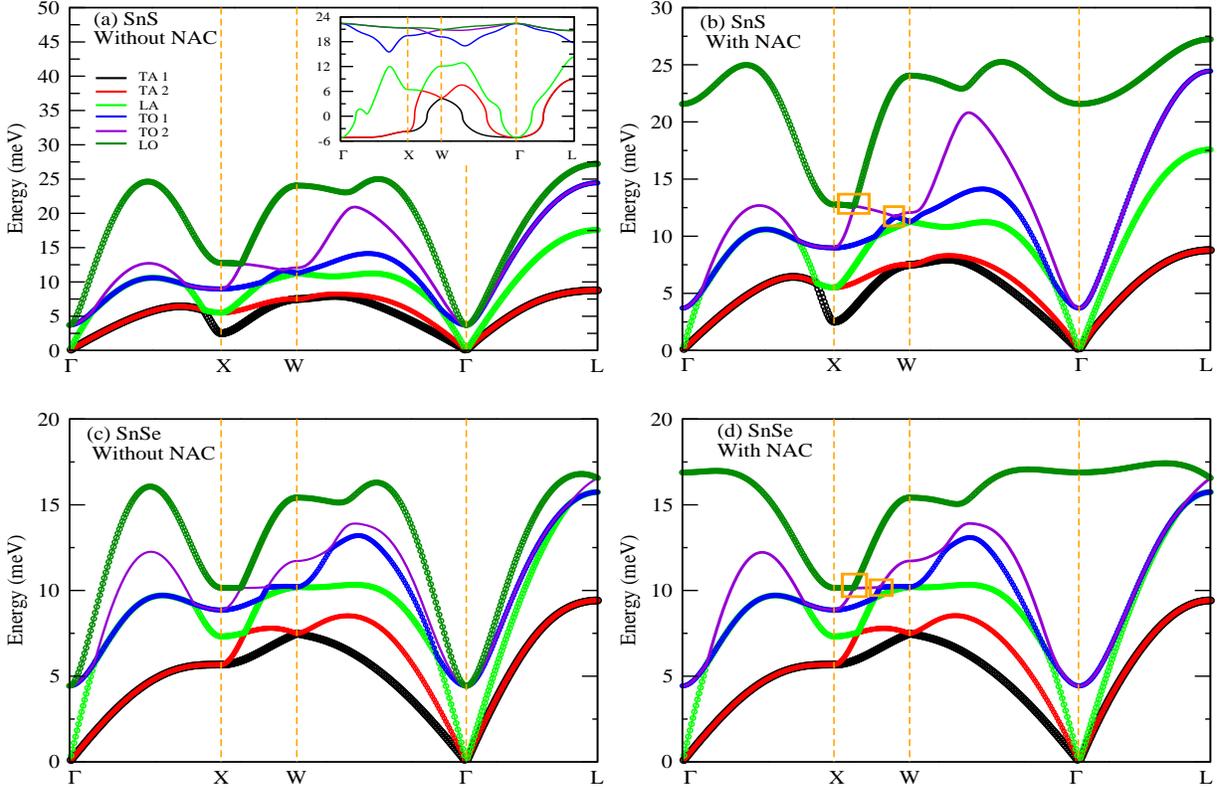} 
    \caption{(Colour online.) Phonon band structure of SnS with (a) not including nonanalytical term correction (NAC) and (b) including NAC. Inset of (a) shows the phonon band structure without considering 4$p$ orbitals as core states. Phonon band structure of SnSe with (c) not including nonanalytical term correction (NAC) and (d) including NAC. Orange boxes are indicating the presence of linear touching of phonon branches. }
    \label{fig:}
  \end{center}
\end{figure*}

  In this work, the structural stability of SnS in rocksalt phase, the phonon \& theromodynamical properties of SnS and SnSe materials with and without NAC and the importance of EPI for getting the topological phase transitions of these two materials have been studied in details. In view of all these properties, a comparative discussion for SnS and SnSe are also described. The phonon band structure of SnS shows no negative frequency when 4$p$ orbitals of Sn atom are considered as valence states for calculating the forces on the atoms. For these two materials, the significance of using NAC within phonon calculations are found for studying the LO-TO splitting, where the thermodynamical properties are also investigated. The estimated values of Debye temperatures ($\Theta_D$) for SnS and SnSe are $\sim$315.0 K and $\sim$201.7 K, respectively. The presence of topological phonon is predicted for these two materials by observing the linear band touching of phonon branches along X-W direction. The decreasing behaviour of bandgap with rise in temperature due to EPI is seen for both the materials, which is showing the bandgap closing at $\sim$700 K and $\sim$900 K for SnS and SnSe, respectively, at L-point. The fundamental bandgap closing is seen at $\sim$700 K for both the materials. The phase transition from topological non-trivial to trivial is observed at this particular temperature for these two materials. The temperature dependent electron and phonon linewidths are discussed together with the lineshift of electron. The linear (non-linear) behaviour is seen form the temperature dependent electron (phonon) linewidth for SnS and SnSe. The momentum resolved spectral function at 100 K and 300 K are also illustrated for both these topological materials.

\begin{figure*}
  \begin{center}
    \includegraphics[width=0.88\linewidth, height=9.5cm]{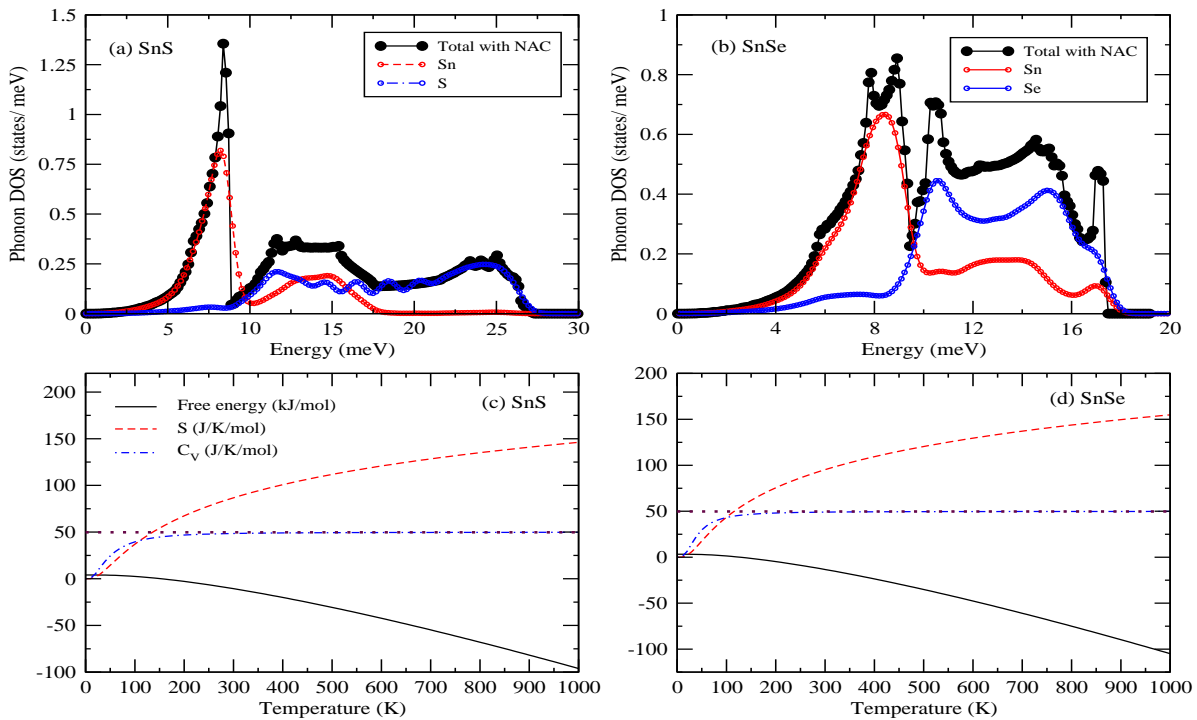} 
    \caption{(Colour online) Phonon total DOS and phonon partial DOS for (a) SnS and (b) SnSe. Specific heat at constant volume ($C_V$), Helmholtz free energy ($F$) and entropy ($S$) for (c) SnS and (d) SnSe. }
    \label{fig:}
  \end{center}
\end{figure*}

\section{Brief description of the theoretical methods}   

  In our present study, we have used the NAC within phonon calculation and EPI through first-principle method using Wannier function. The implementation procedures of these two calculations are briefly described in this section.
  
  The long range Coulomb forces are taken care for any material through the NAC in $ab$-$initio$ phonon calculation. The matrix form of the nonanalytical term is expressed by \cite{togo_nac,giannozzi},
  
\begin{equation}
 \Phi_{\alpha \beta }(\kappa \kappa ^\prime,\textbf{q})=\frac{4\pi e^2}{V_0}\frac{\sum\limits_{\gamma} q_\gamma Z^*_{\kappa,\gamma \alpha} \sum\limits_{\gamma^\prime} q_{\gamma^\prime} Z^*_{\kappa^\prime,\gamma^\prime\beta}}{\sum\limits_{\gamma \gamma^\prime} q_\gamma \epsilon ^\infty _{\gamma \gamma^\prime} q_{\gamma^\prime} }
\end{equation}  
  
  where the volume of the unit cell is denoted by $V_0$, $e$ is the electronic charge, $\textbf{q}$ is the wave vector, the Born effective charge (BEC) tensor for the $\kappa^{th}$ atom is represented by $Z^*_\kappa $ and $\alpha$, $\beta$, $\gamma$, $\gamma^\prime$ are indicating the Cartesian indices. Along with these, tensor of the high frequency static dielectric constant is represented by $\epsilon ^\infty$. It is also noted that this nonanalytical term is a part of harmonic force constants.
  
  In EPW code for particular temperature T, the electron ($\Sigma$) and phonon ($\Pi$) self-energies of n$^{th}$ band and \textbf{k} wavevector are calculated using the following relations \cite{epw,giustino}, 

\begin{eqnarray}
\Sigma_{n\mathbf{k}}(\omega,T)=\sum_{m\nu} \int_{BZ}\frac{d\mathbf{q}}{\Omega_{BZ}}\mid g_{mn,\nu}(\mathbf{k},\mathbf{q})\mid ^{2} \nonumber \\
\times \Bigg\{ \frac{[n_{\mathbf{q} \nu}(T)+f_{m\mathbf{k}+\mathbf{q}}(T)]}{(\omega - (\varepsilon_{m\mathbf{k}+\mathbf{q}}-\varepsilon_{F})+\omega_{\mathbf{q}\nu} + i\delta)} \nonumber \\
+ \frac{[ n_{\mathbf{q} \nu}(T)+ 1 -f_{m\mathbf{k}+\mathbf{q}}(T)]}{(\omega - (\varepsilon_{m\mathbf{k}+\mathbf{q}}-\varepsilon_{F})-\omega_{\mathbf{q}\nu} + i\delta)} \Bigg\} 
\end{eqnarray}

\begin{eqnarray}
\Pi_{\mathbf{q}\nu}(\omega,T)=2 \sum_{mn} \int_{BZ}\frac{d\mathbf{k}}{\Omega_{BZ}}\mid g_{mn,\nu}(\mathbf{k},\mathbf{q})\mid ^{2} \nonumber \\
\times \frac{[f_{n\mathbf{k}}(T)-f_{m\mathbf{k}+\mathbf{q}}(T)]}{\varepsilon_{m\mathbf{k}+\mathbf{q}} - \varepsilon_{n\mathbf{k}} - \omega - i\delta}
\end{eqnarray}

 where, factor 2 is coming due to spin degeneracy and $g_{mn,\nu}(\mathbf{k},\mathbf{q}$) denotes the first-order electron-phonon matrix elements to get the information about the scattering mechanism between two Kohn-Sham states of $m\textbf{k}+\textbf{q}$ and $n\textbf{k}$ associated with a phonon wavevector $\textbf{q}$, which is calculated using DFPT calculation. $\omega_{\mathbf{q}\nu}$ and $\varepsilon_{n\mathbf{k}}$ are representing the phonon frequency of branch index $\nu$ and Kohn-Sham eigenvalue of electron, respectively. $\varepsilon_{F}$ is the Fermi energy. Along with these, $n_{\mathbf{q} \nu}(T)$ and $f_{m\mathbf{k}+\mathbf{q}}(T)$ are denoting the Bose-Einstein and Fermi-Dirac distribution functions, respectively. The real part of electron self-energy is computed by \cite{epw,giustino},

\begin{eqnarray}
\tilde{\Sigma}^{'}_{n\mathbf{k}}(\omega, T) = \Sigma^{'}_{n\mathbf{k}}(\omega, T) - \Sigma^{'}_{n\mathbf{k}}(\omega = \varepsilon_{F}, T)
\end{eqnarray}  
  
 where, $\Sigma^{'}_{n\mathbf{k}}(\omega, T) $ represents the real part of electron self-energy calculated using Eq. 2. Moreover, the imaginary part of electron ($\Sigma^{"}$) and phonon ($\Pi^{"}$) self-energy using EPI within EPW code can be obtained by following the below equations \cite{epw,giustino}, 
  
\begin{eqnarray}
\Sigma^{"}_{n\mathbf{k}}(\omega,T)=\pi \sum_{m\nu} \int_{BZ}\frac{d\mathbf{q}}{\Omega_{BZ}}\mid g_{mn,\nu}(\mathbf{k},\mathbf{q})\mid ^{2} \nonumber \\
\times \lbrace [ n_{\mathbf{q} \nu}(T)+f_{m\mathbf{k}+\mathbf{q}}(T)]\delta(\omega - (\varepsilon_{m\mathbf{k}+\mathbf{q}}-\varepsilon_{F})+\omega_{\mathbf{q}\nu}) \nonumber \\
+ [ n_{\mathbf{q} \nu}(T)+ 1 -f_{m\mathbf{k}+\mathbf{q}}(T)]\delta(\omega - (\varepsilon_{m\mathbf{k}+\mathbf{q}}-\varepsilon_{F})-\omega_{\mathbf{q}\nu}) \rbrace \nonumber \\
\end{eqnarray}
  
 
\begin{eqnarray}
\Pi^{"}_{\mathbf{q}\nu}(\omega,T)=2\pi \sum_{mn} \int_{BZ}\frac{d\mathbf{k}}{\Omega_{BZ}}\mid g_{mn,\nu}(\mathbf{k},\mathbf{q})\mid ^{2} \nonumber \\
\times [f_{n\mathbf{k}}(T)-f_{m\mathbf{k}+\mathbf{q}}(T)]\delta(\varepsilon_{n\mathbf{k}+\mathbf{q}}-\varepsilon_{\mathbf{k}}-\omega)
\end{eqnarray}

  Further, the spectral function ($A_n(\mathbf{k},\omega)$) of electron at temperature $T$ is defined as \cite{epw,giustino,martin}, 

\begin{equation}
A_n{(\mathbf{k},w)} ={\Huge  \frac{1}{\pi} \frac{\mid \Sigma^{"}_{n\mathbf{k}}(\omega) \mid}{[\omega - (\varepsilon^0_\mathbf{k} - \varepsilon_{F}) - \tilde{\Sigma}^{'}_{n\mathbf{k}}(\omega)]^2 + [\Sigma^{"}_{n\mathbf{k}}(\omega)]^2}}
\end{equation}

\section{Computational details}

  For investigating, the phononic structures of SnS and SnSe, the total forces of the atoms are estimated using full-potential linearized-augmented plane-wave (FP-LAPW) based WIEN2k code \cite{wien2k} with considering PBEsol exchange-correlation (XC) functional \cite{pbesol}. The space group of these materials is $F$m-3m, which is utilized in present study along with the lattice parameters 5.753\AA\, and 5.955\AA\, for SnS and SnSe, respectively \cite{asihi_sse}. 5$\times$5$\times$5 k-mesh size is set to perform the present calculation together with fixing the Wyckoff positions at (0.0, 0.0 ,0.0) and (0.5, 0.5, 0.5) for Sn and S (Se), respectively. We have kept the muffin-tin radius 2.2 (2.5) for both Sn and S (Se) in SnS (SnSe) materials. The force convergence criteria is fixed at 0.1 mRy/Bohr for computing the total forces on the atoms using WIEN2k code \cite{wien2k}. R$_{mt}$ $\times$ K$_{max}$ is considered as 8.5 for obtaining the better result. The phonon calculations for these materials are performed using PHONOPY code \cite{phonopy}. Here, finite displacement method is used for phonon calculation with making 2$\times$2$\times$2 supercell. Further, 21$\times$21$\times$21 mesh size is used for calculating phonon density of states (DOS) and thermal properties of the corresponding sample. Moreover, it is noted that spin-orbit coupling (SOC) is not considered in present study because it is not expected to change the motivation of the present work.  
  
  To carry out the calculation of EPI, the EPW \cite{epw} code is chosen, which is interfaced with the Quantum ESPRESSO (QE) \cite{qe} package. Projector augmented wave (PAW) pseudopotential is used both for electronic self-consistent field (scf) calculation and phonon calculation through QE package, where phonon calculation is performed using density functional perturbation theory (DFPT). In this case, PBEsol XC functional is chosen with fixing the cutoff energy at 50 Ry for electronic structure calculation. 10$\times$10$\times$10 and 6$\times$6$\times$6 k-meshes are considered for scf and non-scf (nscf) calculations, respectively, with keeping the convergence threshold for self-consistency at 10$^{-10}$ Ry. Along with this, the phonon calculation in QE is computed with 6$\times$6$\times$6 q-mesh and convergence threshold of 10$^{-14}$ Ry. Afterwards, the self-energy of electron and phonon due to EPI are estimated by utilizing EPW code \cite{epw}. Sn 5$s$, Sn 5$p$, S 3$s$ (Se 4$s$) and S 3$p$ (Se 4$p$) orbitals are chosen as the Wannier orbitals in EPW calculation for SnS (SnSe). The electron and phonon linewidths due to EPI are calculated along W-L-$\Gamma$ direction. In case of self-energy calculation, the value of 10 meV is taken for the Gaussian broadening parameter together with the width of the Fermi surface window of 3.0 eV. In case of EPW calculations, we have fixed the Fermi energy in the middle of the bandgap for both materials. In case of SnS (SnSe), the calculated values of BEC using QE package are found to be $\sim$5.3 ($\sim$5.9) for Sn and $\sim$-5.2 ($\sim$-5.8) for S (Se) and the value of $\epsilon ^\infty$ is estimated to be $\sim$38.41 ($\sim$38.95), which are used both for NAC and EPI calculations.

\section{Results and Discussion} 

\begin{figure*}
  \begin{center}
    \includegraphics[width=0.9\linewidth, height=10.0cm]{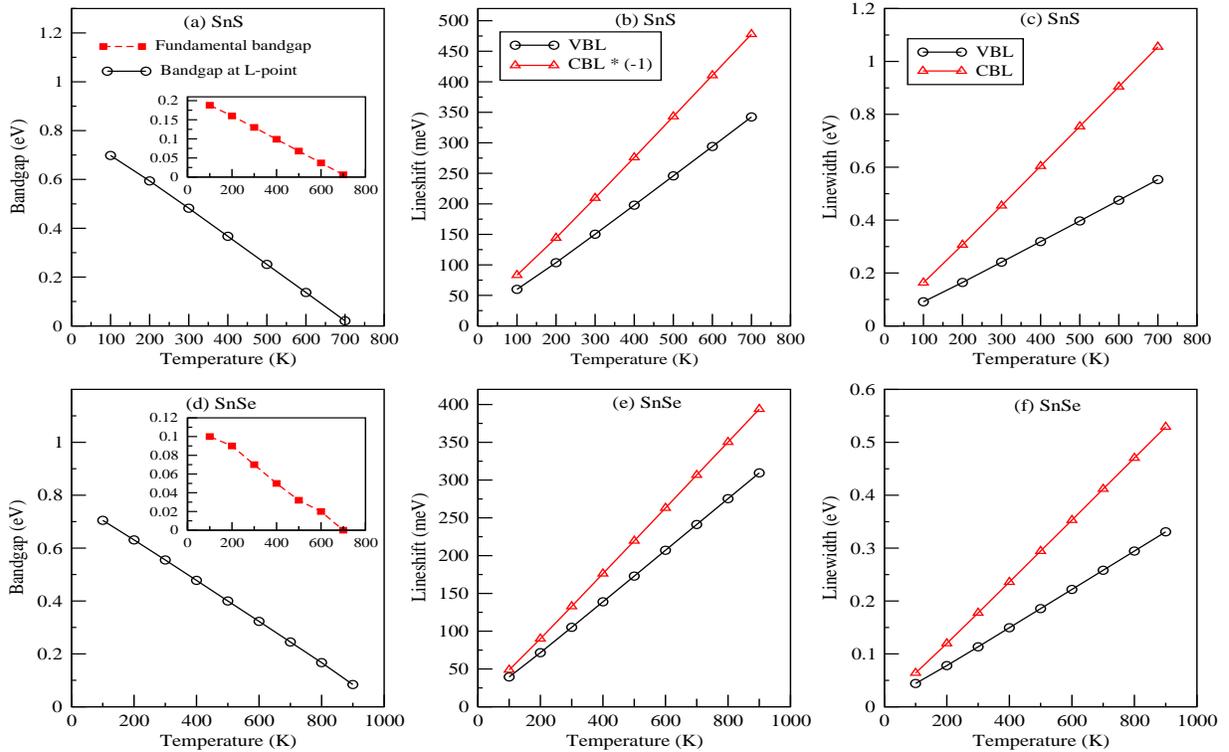} 
    \caption{(Colour online) Calculated values of temperature dependent (a) bandgap at L-point, (b) lineshift and (c) linewidth for VBL and CBL for SnS. Similarly, in the same manner, (d), (e) and (f) are shown for SnSe. The inset of (a) and (b) indicate the temperature dependent fundamental bandgap for SnS and SnSe, respectively.}
    \label{fig:}
  \end{center}
\end{figure*}

\subsection{Phonon and thermodynamic properties}

\begin{figure*}
  \begin{center}
    \includegraphics[width=0.78\linewidth, height=5.0cm]{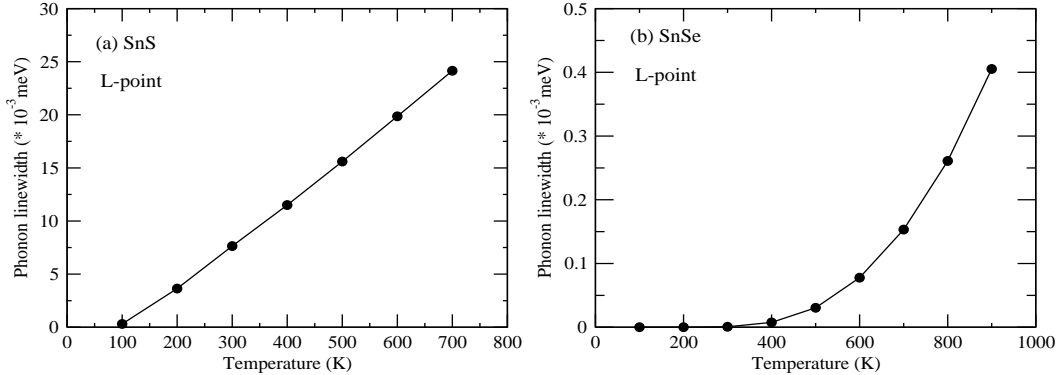} 
    \caption{(Colour online) Calculated values of phonon linewidths as function of temperatures at L-point due to EPI of (a) SnS and (b) SnSe.}
    \label{fig:}
  \end{center}
\end{figure*}

 The phonon band structures of SnS and SnSe in rocksalt phase along the high-symmetric k-directions are shown in Fig. 1. Recently, it is found that including the nonanalytical term correction (NAC) within theoretical phonon calculation is found to be important for describing the experimental phonon band structure of SnTe in rocksalt phase\cite{antik_pla}. Therefore, whether the effect of NAC is also important to describe the phonon dispersion plots of SnS and SnSe or not, we have performed two different phonon calculations with and without including NAC. Six phonon bands are obtained for both materials, where these bands are further divided into two transverse acoustic (TA), one longitudinal acoustic (LA), two transverse optical (TO) and one longitudinal optical (LO) branches as mentioned in the figure. In the inset of Fig. 1(a), the phonon dispersion curve of SnS is showing negative frequency, which follows the previous studies on this compound \cite{skelton_2017, skelton_2021}. Similar to the previous works, it indicates the lattice dynamical instability of SnS in rocksalt phase. But recently, Sk $et.$ $al.$ had shown in their work that the dynamical stability of phonon band structure is strongly dependent on consideration of proper orbitals as core or valence states for computing the forces within the sample using first-principle method \cite{sk_2021}. Therefore, motivated from the aforementioned work, we have considered 4$p$ orbitals of Sn as valence states along with 4$d$, 5$s$ and 5$p$ orbitals for calculating the phonon dispersion curve of SnS, which is shown in Fig. 1(a). These phonon bands are computed without considering NAC, where no negative frequency is observed from Fig. 1(a). Therefore, at this point, it is important to note that SnS in rocksalt phase is dynamically stable in ambient condition and careful experiment is needed for further proof of our work. Two TA branches are degenerate along the observed k-direction except near X-point and W-$\Gamma$ direction. The LA branch of this material is showing degeneracy with one TO (TA) band in $\Gamma$-X direction and around W-point (X-point). The LO phonon branch has shown same energy with one (both) of TO band at around the middle of X-W direction (at $\Gamma$-point). Moreover, after including NAC within phonon calculation, the degeneracy of LO and TO bands at $\Gamma$-point is lifted, which is clearly shown from Fig. 1(b). The energy difference due to LO-TO degeneracy lifting for SnS is found to be $\sim$17.7 eV after including NAC. In case of other bands, they are showing similar behaviour as described in earlier case for SnS material. It is noted that further calculations of phonon density of states (DOS) and thermal properties of this material are carried out using these phonon branches, where NAC and 4$p$ orbitals of Sn are considered for performing the phonon calculation. Now, the Fig. 1(c) illustrates the phonon dispersion curve of SnSe without including NAC. The overall phonon band features of SnSe are showing almost similar behaviour as found for SnS from Fig. 1(a) except around the L(X)-point. In case of SnSe at L(X)-point, the LA branch is (not) showing degeneracy with one of TO (TA) branch and the another TO branch is degenerate with the LO branch. Moreover, after including NAC, the degeneracy between LO and TO branches is lifted as seen from Fig. 1(d), which is typically seen after considering NAC in phonon calculation. The energy difference between the L0 and TO phonon branches are estimated to be $\sim$12.4 eV after considering NAC. For SnS (SnSe), the highest phonon band energy is found to be $\sim$27.2 ($\sim$16.9) meV and the estimated energy difference between before and after including NAC is $\sim$20.3 ($\sim$10.98) meV at $\Gamma$-point. It is noted that the lattice constant together with mass of X atom increases when one goes from SnS to SnSe in case of SnX (X= S, Se) materials. Therefore, it may be the reason for observing the decreasing trend in phonon frequency on moving from SnS to SnSe. It is clearly seen from Figs. 1(b) and 1(d) that along X-W direction for both the materials, the optical branches are showing the presence of linear band crossing (marked by orange boxes in figures). In case of bosonic system as compared to fermionic system, the presence of linear band crossing along with the band inversion usually illustrates the presence of different kind of topological Weyl or Dirac phonons \cite{chen_2021,liu_weylphn}. Moreover, type II Weyl phonons are generally observed when the tilted linear band crossing is seen in phonon band structure. CdTe is the first realistic material to categorize as type II Weyl phonon in a cubic system. In case of CdTe, the existence of topological phonon is found along X-W direction. Therefore, the linear band touching of the phonon branches along X-W direction for SnS and SnSe may indicate the presence of topological phonons, which is also recently observed for ZnSe in Wurtzite structure \cite{liu_weylphn}. Thus, chalcogenide based materials may be played important role for obtaining the realistic topological phonon. Moreover, the behaviour of topological phonon for SnS and SnSe needs to be investigated in details. For SnSe, the further studies of thermodynamic and phonon DOS related with lattice vibrations are performed with these phonon modes where NAC is considered in calculation. 
  
  The calculated phonon total DOS (TDOS) and partial DOS (PDOS) for SnS and SnSe materials are shown in Figs. 2(a) and 2(b), respectively. For SnS (SnSe) in phonon TDOS, the position of first peak is found at $\sim$8.4 meV, where the major contribution for this particular peak is coming from the Sn atom. The next broadened peak, which is found at 11.5 meV - 15.7 meV (10.3 meV - 17.2 meV), is seen in phonon TDOS due to the mixed contributions of Sn and S (Se) atoms for SnS (SnSe) material. In case of SnS, one additional peak is found at $\sim$24.3 meV due to S atom, whereas this peak is not present for SnSe material due to Se atom. It is known that the atomic mass of Sn is more closer to Se than S. This may be the reason to obtain three major peaks for SnS, whereas only two major peaks are seen for SnSe. It is seen that the behaviour of phonon TDOS of SnSe has nicely followed the phonon TDOS of SnTe \cite{antik_pla}. Further, the computed phonon TDOS of SnSe is nicely matched with the previously reported theoretical data \cite{skelton_2021, zhou_2020}. 
  
\begin{figure*}[]
   \begin{subfigure}{0.45\linewidth}
   \includegraphics[width=0.92\linewidth, height=6.0cm]{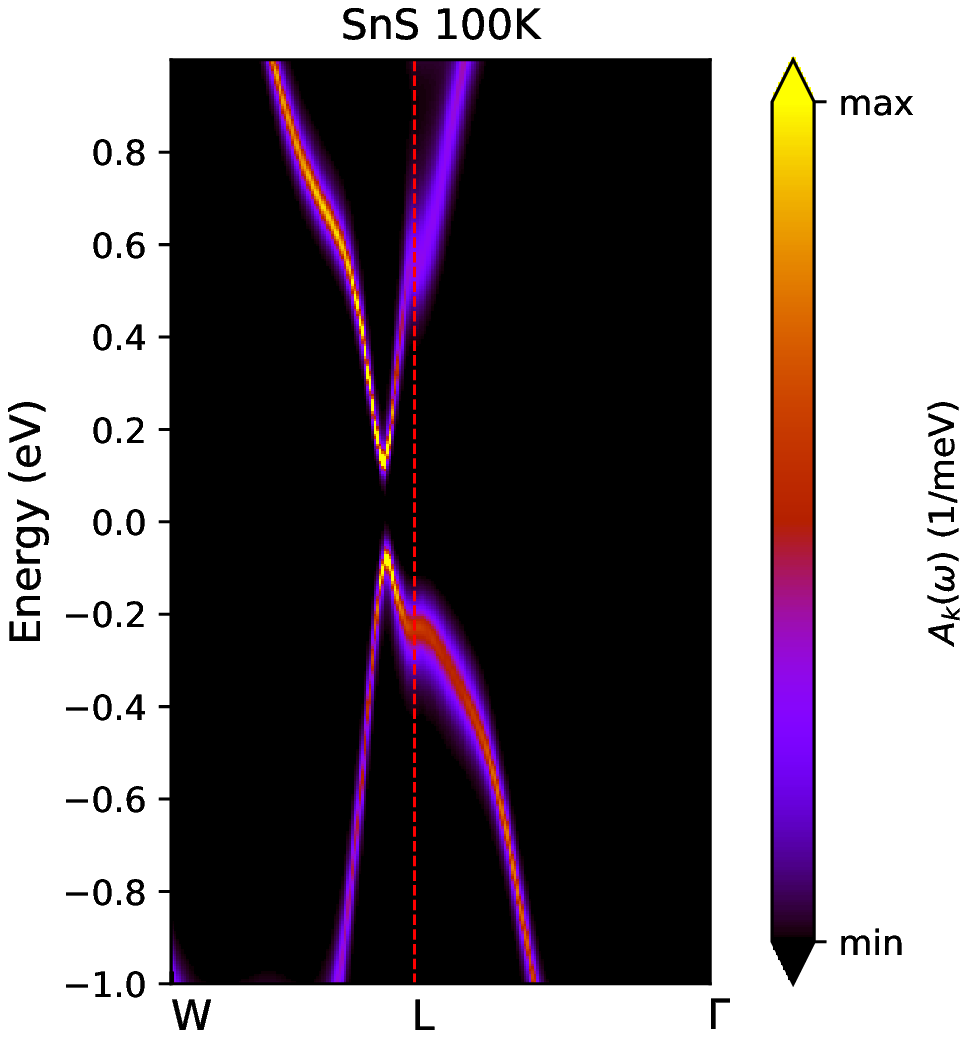}
   \caption{}
   \label{fig:} 
\end{subfigure}
\begin{subfigure}{0.45\linewidth}
   \includegraphics[width=0.92\linewidth, height=6.0cm]{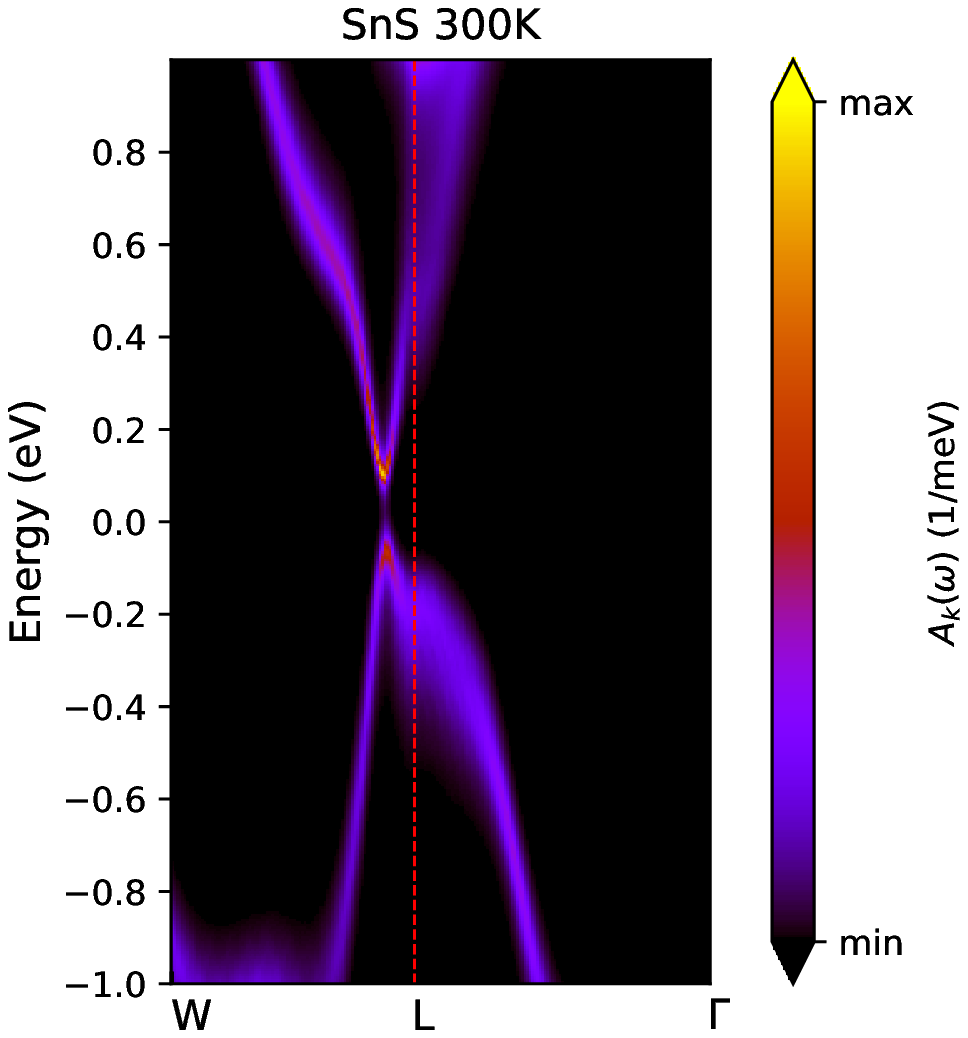}
   \caption{}
   \label{fig:}
\end{subfigure}
\caption{(Colour online) Momentum-resolved spectral functions of SnS at (a) 100 K and (b) 300 K along W-L-$\Gamma$ direction.}
\end{figure*}

\begin{figure*}[]
   \begin{subfigure}{0.45\linewidth}
   \includegraphics[width=0.92\linewidth, height=6.0cm]{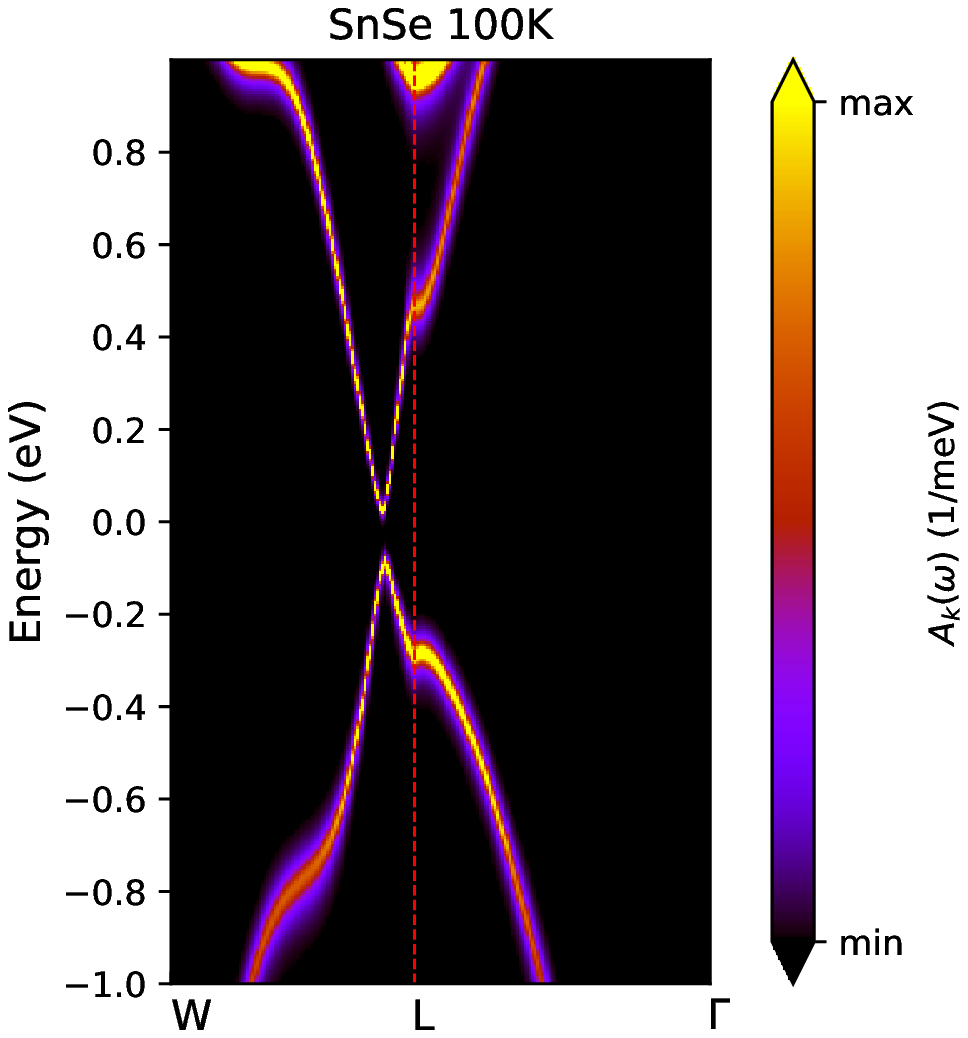}
   \caption{}
   \label{fig:} 
\end{subfigure}
\begin{subfigure}{0.45\linewidth}
   \includegraphics[width=0.92\linewidth, height=6.0cm]{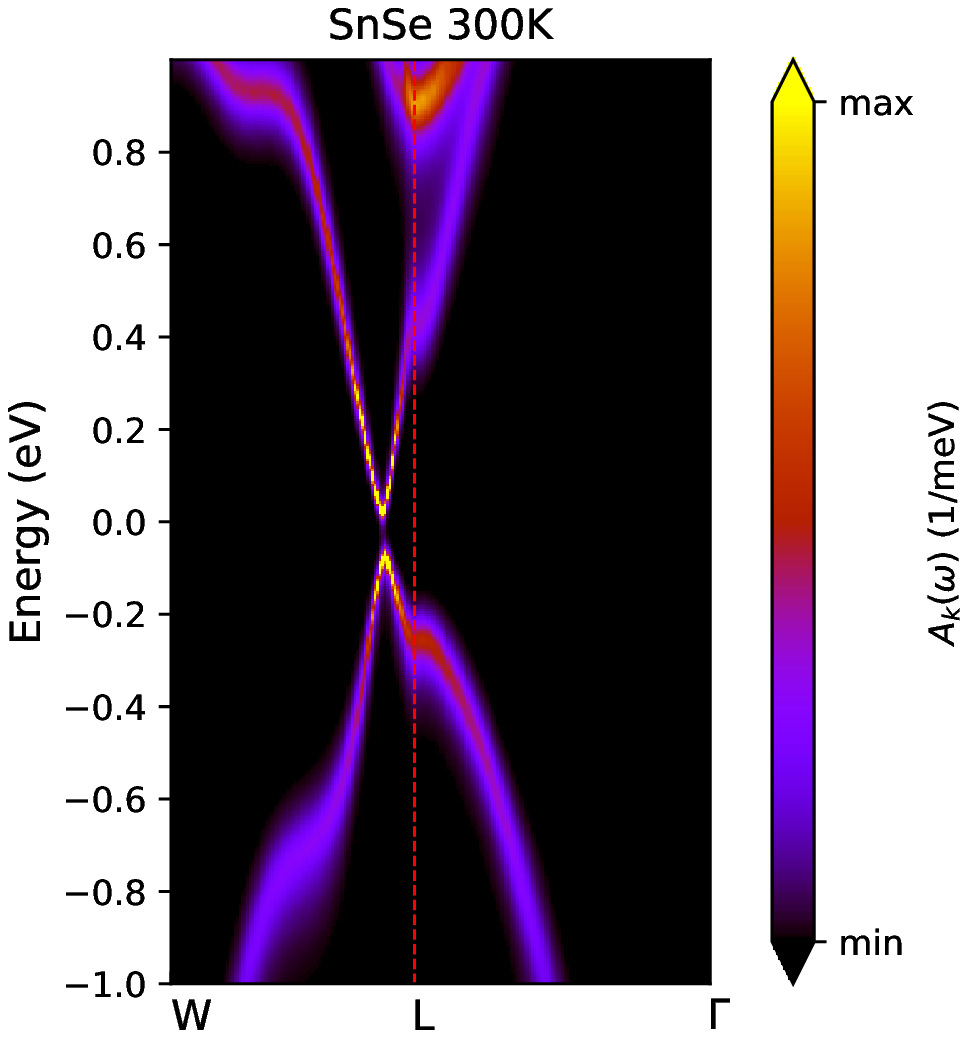}
   \caption{}
   \label{fig:}
\end{subfigure}
\caption{(Colour online) Momentum-resolved spectral functions of SnSe at (a) 100 K and (b) 300 K along W-L-$\Gamma$ direction.}
\end{figure*}  

 The different thermodynamical quantities of SnS and SnSe are calculated for the temperature range of 0 K to 1000 K and plotted in Figs. 2(c) and 2(d), respectively. The thermodynamic relations, which are dependent on the phonon vibrations of sample, used to calculate the values of specific heat at constant volume ($C_V$), Helmholtz free energy ($F$) and entropy ($S$) are given as \cite{phonopy}, 
 
\begin{equation}
C_V={\sum\limits_{\textbf{q}j}}k_B\bigg(\frac{\hbar\omega_{\textbf{q}j}}{k_BT}\bigg)^2\frac{exp(\hbar\omega_{\textbf{q}j}/k_BT)}{[exp(\hbar\omega_{\textbf{q}j}/k_BT)-1]^2}
\end{equation}

\begin{equation}
 F=-k_BTlnZ
\end{equation}

where partition function ($Z$) denotes as,

\begin{equation}
Z=exp(-\phi/k_BT) \prod\limits_{{\textbf{q}j}}\frac{exp(-\hbar\omega_{\textbf{q}j}/2k_BT)}{[1-exp(-\hbar\omega_{\textbf{q}j}/k_BT)]}\nonumber
\end{equation}

\begin{equation}
 S = -\frac{\partial F}{\partial T}
\end{equation}

where $\phi$, $k_B$, $\hbar$ are the crystal potential energy, the Boltzmann constant and the reduced Planck's constant, respectively. Also, $\omega_{\textbf{q}j}$ represents the phonon frequency for mode of \textbf{q}, j and $T$, where $T$ denotes the temperature in absolute scale. It is known that at higher temperature, the value of $C_V$ reaches the Dulong-Petit limit, which is found to be $\sim$49.8 J/K/mol for both the materials. In the figures, it is denoted by the dotted line. In case of SnS, at $\sim$350 K, the value of $C_V$ is seen to be equal to Dulong-Petit limit, whereas for SnSe, the temperature value is $\sim$330 K. The phonon dispersion curve of these two materials show higher values of phonon energy (or frequency) for SnS than SnSe, which may be one of the possible reason to obtain the temperatures decreasing behaviour for high temperature specific limiting case. The temperature independent behaviour is typically seen above the temperature estimated by Dulong-Petit law. But, in case of lower temperature region, generally $C_V$ is proportional to $T^3$ as evident from Debye model. Next, the values of $S$ are showing the increasing nature with rise in temperature for both the materials, where the value of $S$ at 300 K for SnS (SnSe) is found to be $\sim$86.7 ($\sim$95.2) J/K/mol. In case of DFT based electronic structure calculation, generally we have considered the static lattice model, where the Born-Oppenheimer approximation is usually applied for computing the ground state energy eigenvalues of any material. But, the presence of quantum behaviour suggests the existence of lattice vibration even at zero temperature. Therefore, in order to estimate the zero-point energy of these two materials, the temperature dependent $F$ is calculated. The estimated value of zero-point energy for SnS (SnSe) is $\sim$4.0 ($\sim$3.2) J/K/mol. The reason of decrements of zero-point energy value is due to the presence of higher and lower frequency of SnS and SnSe, respectively, which is already discussed above. The value of $F$ becomes negative after 150 K (130 K) for SnS (SnSe). Further, the Debye temperature ($\Theta_D$) defines the transition temperature above which all the phonon branches have obtained sufficient thermal energy for vibrations \cite{ashcroft}. One of the way to find out the value of $\Theta_D$ from first-principle calculation is taking the maximum phonon frequency ($\omega_{max}$) as Debye frequency \cite{ashcroft}. Here, following this mentioned way, the computed values of $\Theta_D$ are $\sim$315.0 K and $\sim$201.7 K for SnS and SnSe, respectively. Moreover, the effect of temperature on the interaction between fermions and bosons (for example electron-phonon interaction) always brings new insight of any material.

\subsection{Temperature-induced electron-phonon interaction} 

  In order to get the transition from non-trivial topological insulating phase to normal insulating phase, the bandgap's closing with respect to temperature is needed. It is found form the literature that EPI plays important role for SnTe in rocksalt structure to show the mentioned phase transition \cite{flores}. Therefore, studying the topological phase transitions for other famous TCIs of SnS and SnSe are significant through considering the temperature dependent EPI within first-principle calculations.  Moreover, it is studied by Sihi $et.$ $al.$ \cite{asihi_sse} that PBEsol is failed to provide the semiconducting bandgap in case of SnS, but mBJ and $G_0W_0$ methods have shown the bandgap. However, the behaviour of the bands are almost similar for PBEsol, mBJ and $G_0W_0$ methods. Therefore in present case, for obtaining more realistic semiconducting dispersion curve of SnS from PBEsol functional, a rigid shifting for VB (CB) towards lower (higher) energy with respect to $E_F$ is given. It is known that the bandgap of these two materials are found near the high-symmetric L-point \cite{asihi_sse}. Thus, we have focused to study the effect of EPI at L-point of valence band maxima (VBL) and conduction band minima (CBL). The temperature dependent bandgap values due to EPI are plotted in Fig. 3(a) and Fig. 3(d) for SnS and SnSe, respectively. The bandgap is calculated by considering the Kohn-Sham energy difference of VBL and CBL plus the energy difference of lineshifts ($i.e.$ real part of electronic self-energy due to electron-phonon interaction ($Re\Sigma$ ($\omega$)) corresponding to these two points, which is defined as E$_{bandgap}$ = (E$_{CBL}$ - $Re \Sigma_{CBL}$) - (E$_{VBL}$ - $Re \Sigma_{VBL}$) at L-point. The Figs. 3(a) and 3(d) show monotonically decreasing behaviour of the bandgap values with increasing temperatures for both materials. The value of bandgap at 300 K is found to be $\sim$0.48 ($\sim$0.56) eV for SnS (SnSe). The bandgaps of SnS are found to have lower values than SnSe for the studied temperature range. This is showing opposite behaviour of temperature dependent bandgaps due to EEI for both these materials \cite{asihi_sse}. Therefore, the bonding and anti-bonding strength between Sn 5$p$ - S 3$p$ orbitals are becoming more stronger (weaker) than Sn 5$p$ - Se 4$p$ orbitals at L-point, when EEI (EPI) is considered. The bandgap's value at L-point becomes almost zero at $\sim$700 K ($\sim$900 K) for SnS (SnSe). In present scenario, it is observed that the electronic states are moving downward (upward) through the phonon mediated coupling with higher (lower) energy electronic states. For SnTe, Querales-Flores $et.$ $al.$ had discussed in their work that the EPI near the L-point depends on the phonons which are seen close to $\Gamma$ and X-points \cite{flores}. Therefore, the similar discussion is also valid for SnS and SnSe materials, which have the same crystal structures. But, SnS shows higher phonon energy than SnSe at both $\Gamma$ and X-points as evident from Figs. 1(b) and 1(d). This behaviour suggests the presence of more stronger electron-phonon coupling in case of SnS than SnSe. This may be the reason for getting lower transition temperature for SnS as compared to SnSe. Moreover, it is known that the band inversion is seen near the L-point for these two samples \cite{y_sun}. Here, the shifting of VBL (CBL) in upward (downward) direction is suggesting that the band inversion is removed when the bandgap closes at L-point. In addition to this, the temperature dependent fundamental bandgap are also plotted in the insets of Figs. 3(a) and 3(d) for SnS and SnSe, respectively. In this case, the bandgap is closing at temperature $\sim$700 K for both the materials. Therefore, the transition temperature at which the non-trivial topological phase changes to normal insulator is estimated to be $\sim$700 K for SnS and SnSe. Furthermore, this non-linear temperature dependent bandgaps for both materials have drawn the attention to discuss further temperature dependent lineshift and linewidth of electronic states due to EPI. 

  The computed values of lineshifts for VBL and CBL are shown in Figs. 3(b) and 3(e) for SnS and SnSe, respectively, because the band inversion is more profoundly observed around the L-point for both the compounds. In case of these two materials, the values of lineshifts are linearly increasing (decreasing) with rise in temperature for VBL (CBL) of SnS and SnSe. The estimated values of lineshifts for SnS have larger (smaller) value than SnSe at any particular temperature for VBL (CBL). It is a direct consequence to evident the presence of stronger electron-phonon coupling for SnS than SnSe in the observed temperature range. The values of lineshifts for SnS and SnSe at temperature 300 K for VBL (CBL) are estimated to be $\sim$150.3 ($\sim$-209.6) meV and $\sim$105.1 ($\sim$-132.9) meV, respectively. The opposite behaviour of lineshifting for VBL and CBL is responsible for bandgap closing at transition temperature for these materials. Moreover, above the transition temperature, this lineshift of the electronic states are also responsible to remove the behaviour of band inversion as observed at VBL and CBL for these materials. This is a direct evidence of transition from non-trivial to trivial topological phase for any material. Now to understand the non-linear temperature dependent bandgaps of these materials, which are seen to be similar with SnTe \cite{flores}, we have studied the temperature dependent linewidths of both electron and phonon quasiparticles due to the EPI.
  
\begin{figure*}
  \begin{center}
    \includegraphics[width=0.8\linewidth, height=5.5cm]{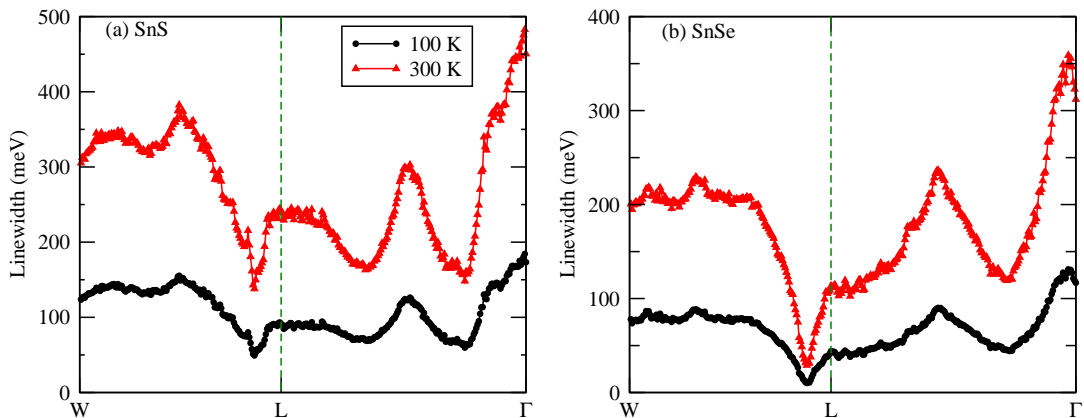} 
    \caption{(Colour online) Calculated values of linewidth due to EPI along W-L-$\Gamma$ direction at 100 K and 300 K for the top most VB of (a) SnS and (b) SnSe.}
    \label{fig:}
  \end{center}
\end{figure*}   

  The linewidth, which is equal to the twice of imaginary part of electronic self-energy ($Im\Sigma$ ($\omega$)) due to EPI, provides the insight of many-body interaction's effect between two quasiparticles. In addition to this, the inverse of linewidth gives the information of relaxation time ($\tau$) of quasiparticle state, where it is known that understanding the behaviour of $\tau$ is quite significant for studying the transport properties of any material. The calculated values of linewidths of VBL (CBL) are shown in Figs. 3(c) and 3(f) for SnS and SnSe, respectively. It is clearly seen from the figures that in case of both materials, the values of linewidths are showing linearly increasing behaviour with rise in temperatures at VBL and CBL. The calculated values of linewidths of VBL (CBL) at temperature 100 K, 300 K, 500 K and 700 K are found to be $\sim$0.09 ($\sim$0.16) eV, $\sim$0.24 ($\sim$0.46) eV, $\sim$0.40 ($\sim$0.75) eV and $\sim$0.55 ($\sim$1.06) eV for SnS and $\sim$0.04 ($\sim$0.06) eV, $\sim$0.11 ($\sim$0.18) eV, $\sim$0.19 ($\sim$0.29) eV and $\sim$0.26 ($\sim$0.41) eV for SnSe, respectively. This increasing trend is found to be similar with SnS and SnSe materials, where the temperature dependent $Im\Sigma$ ($\omega$) of SnS and SnSe are computed by considering the EEI \cite{asihi_sse}. But, the non-linear linewidths for VBL and CBL due to EPI are described for SnTe, which is not similar with our present case \cite{flores}. The masses of Sn and Te atoms are comparable with each other but the mass of Sn atom is much higher than the mass of Se (S) atom for SnSe (SnS) material. Hence, phonon frequencies of SnS (SnSe) are much higher than SnTe. Also, in case of SnTe, both 5$p$ orbitals of Sn and Te ions are responsible for forming the band at these two k-points ($i.e.$ VBL \& CBL), but for SnS (SnSe), Sn 5$p$ and S 3$p$ (Se 4$p$) orbitals are contributing at VBL \& CBL. It is known from the basic knowledge of atomic physics that the spread of 5$p$ orbitals are more than 4$p$ (3$p$) orbitals of any material. Therefore for SnTe, the interaction between the electrons of two 5$p$ orbitals associated with phonon modes has provided non-linear behaviour in temperature dependent linewidth's plot, whereas the interaction due to electrons of 5$p$ and 3$p$ (4$p$) orbitals with the corresponding phonon branches give linear trends with rise in temperatures for SnS (SnSe). The magnitude of linewidth's values for CBL are seen to be more than VBL for both the materials. The rate of increment of linewidths with respect to temperature is found to be higher for SnS than SnSe, where the estimated values of slope for VBL and CBL are found to be $\sim$7.8 ($\sim$15.0) $\times$ 10$^{-4}$ eV/K and $\sim$3.6 ($\sim$5.83) $\times$ 10$^{-4}$ eV/K for SnS (SnSe), respectively. It is showing such behaviour because the corresponding phonon branches of SnS material have higher energy than the SnSe. Therefore, in case of SnS, the availability of phase space for scattering of these quasiparticles are much higher than SnSe. However, the reason of getting the non-linearity in temperature dependent bandgap data is not possible to describe from the linear behaviour of linewidths and lineshifts of electronic states for both these materials. Thus, the temperatures dependent phonon linewidths due to EPI are also discussed for these two materials. The Figs. 4(a) and 4(b) illustrate the phonon linewidths as function of temperature for SnS and SnSe, respectively, at L-point. Here, the phonon linewidth is calculated by adding the linewidths of all phonon branches at particular temperature for L-point. The comparison between the Figs. 4(a) and 4(b) provides the evidence of higher values of phonon linewidth for SnS than SnSe at particular temperature. At L-point for both the materials, it is clearly seen from these figures that the non-linear behaviour is observed for phonon linewidth with rise in temperature. These non-linear nature of phonon linewidth may be responsible for closing the bandgap and changing the band inversion order near to Fermi level for SnS and SnSe. Moreover, in order to see the presence of coherent and incoherent features of the electronic states near to the Fermi level, we have calculated the momentum-resolved spectral function at different temperatures for both the samples.


 Figs 5(a) (6(a)) and 5(b) (6(b)) show the momentum-resolved spectral function of SnS (SnSe) at 100 K and 300 K, respectively, along W-L-$\Gamma$ direction. The yellowish (violetish blue) colour in these figures indicates the coherent (incoherent) part of the spectrum due to EPI. It is clearly seen from these figures that with rise in temperature the incoherent features of the spectrum are increasing for both the materials, which is a common behaviour typically seen with rise in temperature. Around the L-point, the presence of incoherent states are found to be more populated in SnS than SnSe material. This behaviour suggests the presence of more scattering point in SnS around L-point than SnSe, which will be responsible for getting faster rate of decreasing of bandgap with temperature for SnS as compared to SnSe. It is also noted that the change of bandgap with rise in temperature is directly seen from these figures for both the materials. These predicted results can be verified through angle-resolved photoemission spectroscopy (ARPES) data.      

%
%
       
  In addition to these, for further description of many-body interaction on the electronic quasiparticle states due to EPI, the linewidths of top most VB are computed along W-L-$\Gamma$ direction at 100 K and 300 K for SnS and SnSe, where the calculated results are shown in Figs. 7(a) and 7(b), respectively. These calculated values will be helpful for comparing the data with ARPES experiment in future for SnS and SnSe compounds. In overall, the linewidths of SnS for this particular direction are showing higher values than SnSe at a fixed temperature. It is noted that at 100 K, the value of linewidth for SnS (SnSe) is found to be $\sim$49.3 ($\sim$10.2) meV at the particular k-point where the fundamental bandgap is seen from the corresponding electronic dispersion curve \cite{asihi_sse}. At this aforementioned k-point, the linewidth is showing much lower value than the other observed k-point along W-L-$\Gamma$ direction at 100 K for SnS and SnSe. This behaviour indicates that the lifetime of the electronic states will be higher at this particular k-point than the other studied k-points for both materials. It is also noted that the difference between values of linewidths of 100 K and 300 K is not same for all the observed k-points along W-L-$\Gamma$ direction. It suggests that the phonon mediated interaction with electrons is showing different strength at different k-points for particular material. The linewidths of SnS are showing much differed values than SnSe, when one moves from temperature 100 K to 300 K. This behaviour is evident for the presence of stronger EPI in SnS than SnSe material, which may also have effect on the transport and other physical properties along with topological behaviour of these two materials. The values of linewidths at W, L and $\Gamma$-points for SnS (SnSe) are found to be $\sim$305.3 ($\sim$200.7) meV, $\sim$241.1 ($\sim$113.4) meV and $\sim$450.4 ($\sim$312.0) meV, respectively at temperature 300 K. It is also noted that the values of linewidths of electronic quasiparticle due to EPI for these two materials at 100 K is found to be almost comparable with the values of linewidths of electronic quasiparticle due to EEI at 700 K along W-L-$\Gamma$ direction, except around the L-point for SnSe \cite{asihi_sse}.  
  

\section{Conclusions} 

  Here, the importance of nonanalytical term correction (NAC) for understanding the phonon band structure along with the thermodynamical properties and the insight of temperature induced EPI are focused to investigate for the rocksalt phase of SnS and SnSe materials. The phase stability of rocksalt SnS in ambient condition is found after considering Sn 4$p$ orbitals as valence electrons, whereas the previous theoretical works have claimed the phase instability of this material. The LO-TO splitting together with the importance of NAC is discussed for both the materials. The different thermodynamical properties are described for these two materials and the value of Debye temperatures for SnS (SnSe) is estimated to be $\sim$315.0 K ($\sim$201.7 K). The phonon band structure of SnS and SnSe have shown linear touching in several \textbf{q}-points along X-W direction, which may be indicating the presence of topological phonon along this direction. Moreover, the temperature dependent bandgap due to EPI shows non-linear behaviour for both the materials. The non-linearity of bandgap is explored with investigating the linewidths and lineshift of valence band (VB) and conduction band (CB) at L-point due to EPI. The values of linewidths for VB and CB at L-point for 300 K are estimated to be $\sim$0.24 ($\sim$0.11) eV and $\sim$0.46 ($\sim$0.18) eV for SnS (SnSe), respectively. The topological phase transition is found at $\sim$700 K for SnS (SnSe) via closing the bandgap along with removing the band inversion. The present study on SnS and SnSe opens up the possibility of using these materials for device making application due to the behaviour of temperature induced topological phase change. All these results discussed in this work can be verified by performing the photoemission spectroscopy and neutron inelastic scattering experiments.

\end{document}